\documentclass[fleqn,showpacs,preprint,preprintnumbers,amsmath,amssymb]{revtex4}

\usepackage{graphicx}
\usepackage{dcolumn}
\usepackage{bm}

\begin{document}

\title{Universal characteristics of ion-acoustic wave dynamics in magnetized plasmas with emphasis on Tsallis distribution}

\author{M. Akbari-Moghanjoughi}
\affiliation{Azarbaijan University of
Tarbiat Moallem, Faculty of Sciences,
Department of physics, 51745-406, Tabriz, Iran}

\date{\today}

\begin{abstract}
Using the extended Poincar\'{e}-Lighthill-Kuo (PLK) reductive perturbation method, which incorporates the phase-shift variations, it is shown that common features on propagation and head-on collisions of ion-acoustic waves exist for a magnetized plasmas of different inertial-less particle distributions. For instance it is remarked that, the soliton amplitude is always independent of magnetic field strength while strictly depends on its angle regarding the propagation direction. Both types of solitons (compressive or rarefactive) are shown to exist which are defined through the critical angle $\gamma=\pi/2$ or other critical values depending on plasma fractional parameters. These critical plasma parameter values also define the sign of head-on collision phase shift. Furthermore, it is proved that for a given set of plasma parameters there is always a relative angle of propagation regarding to that of the magnetic-field for which the soliton width is maximum. Current findings applies to a wide range of magnetized plasmas including those containing background dust ingredients or two-temperature inertial-less particles and may be used to study laboratory or astrophysical magnetoplasmas.
\end{abstract}


\maketitle

\section{Introduction}\label{intro}

Historically, the first experimental evidence of the ion-acoustic solitary excitations in plasma has been established by Ikezi et al. in 1970 \cite{ikezi}. However, due to the wide and inevitable applicability in fast growing plasma technology and astrophysical sciences, the theoretical prediction of nonlinear density excitations in ionized environments has been taken place much earlier \cite{vedenov, sagdeev}. Since then, several methods such a pseudopotential and reductive perturbation techniques \cite{davidson} has been employed to explore interesting properties of such waves, in particular, electron-acoustic, ion-acoustic or dust-acoustic nonlinear structures in both classical and quantum plasmas with different species and distributions \cite{Surko, Rizzato, Nejoh, Berezh1, Berezh2, Berezh3, Berezh4, Popel, Shukla2, Saleem, Yu, Mushtaq, Tiwari1, Tiwari2, Mamun1, Mamun2, Mamun3, Mamun4, Anowar, nejoh, mahmood1, mahmood2, mahmood3, abdolsalam, akbari1, akbari2}. More recently, using magnetohydrodynamics (MHD) model, it has been shown that despite the differences in species type or equations of state (EoS), essential dynamics of nonlinear excitations in plasmas in the presence of an ambient magnetic field possess common features \cite{akbari3, akbari4}. During the past decades, however, a new motivation has started towards the study of such nonlinear structures in positron containing plasmas due to their presence in Van Allen radiation belts and near the polar cap of fast rotating neutron stars \cite{Mic}, active galactic nuclei \cite{Mil}, quasars and pulsar magnetosphere \cite{Ray, Mic2} and solar atmosphere \cite{Tand}. It is also strongly believed that the electron-positron plasma excitations may be the key to understanding of the evolution of early Universe \cite{Rees, Misner}.

In recent years many attention have been payed to the generalized Boltzmann-Gibbs thermodynamics first formulated by Tsallis \cite{tsallis}. It has been proposed that, this generalization of particle distributions, for instance, may arise due to transfer of the plasma particles through the strongly turbulent, non-integer dimensional, and irregular media which may be encountered in astrophysical situations \cite{tsallis2}. This anomalous irregularities in space plasmas may also be due to the effects of unknown external forces acting in astrophysical environments or due to the wave-particle interaction leading to Tsallis-like distribution of plasma species causing a high-energy tail to appear in the distribution function of the particles. However, the origin of high-energy superthermal (non-Maxwellian) charged particles, observed in solar wind, magnetosphere, interstellar medium and auroral zone \cite{mendis, lazar}, is one of the unsolved problems in the field of space and astrophysical plasmas. The application of Tsallis-like (Kappa) distribution in modeling the space plasmas had been first suggested long before the discovery of Tsallis thermodynamics by Vasyliunas in 1968 \cite{Vas}, and was later adopted by many authors in various physical contexts. There has been extensive theoretical work focused on the effects of superthermal particles on different types of linear and nonlinear collective processes in plasmas \cite{dub, Milo, Chat, Gun, Leu, Jun, Leu2, Trib, Dum, Rab, Rub, Ait, Abb, Na}. Among other applications of the superthermal q-nonextensive electron distributions are the interpretation of observations in the Earth foreshock and solar wind models with coronal electrons first observed in various experimental plasmas, such as laser-matter interactions or plasma turbulence \cite{Mag}. Unfortunately, however, there is a lack of investigation of such plasma excitations in the presence of magnetic field which is inevitable in astrophysical situations.

In current investigation we attempt to generalize the previous studies on propagation and collision of ion-acoustic solitary waves in three-component electron-positron-ion magnetoplasmas employed for various electron-positron distributions to find common features ruling the dynamics of such nonlinear structures. We further pay special focus on superthermal distribution which is most relevant to space plasmas. The article is organized in the following manner. The basic normalized hydrodynamics equations are introduced in section \ref{basic}. Nonlinear evolution and collision parameters are derived in section \ref{shift}. Global and specific features are presented in Secs. \ref{discussion1} and \ref{discussion2}. The final remarks are drawn in section \ref{conclusion}.

\section{Hydrodynamic Model Equations}\label{basic}

We consider a three-component uniformly magnetized plasma consisting of inertial positive hot species, say ions and inertial-less negative and positive ingredients, say electrons and positrons the masses of which is ignored compared to inertial components. Furthermore, we assume inertial particle temperature to be much lower compared to that of inertial-less ones, i.e., $T_{i}\ll T_{+,-}$. The complete set of three-dimensional normalized MHD equations, may be written as
\begin{equation}\label{normal}
\begin{array}{l}
\frac{{\partial {n_i}}}{{\partial t}} + \nabla \cdot({n_i}{{\bf{u}}_i}) = 0,\hspace{3mm}{{\bf{u}}_i} = {\bf{i}}{u_i} + {\bf{j}}{v_i} + {\bf{k}}{w_i}, \\
\frac{{\partial {{\bf{u}}_i}}}{{\partial t}} + ({{\bf{u}}_i}\cdot\nabla ){{\bf{u}}_i} + \nabla \varphi  + \frac{\sigma }{{{n_i}}}\nabla {n_i} + \bar \omega ({{\bf{u}}_i} \times {\bf{k}}) = 0, \\
{\nabla ^2}\varphi  = ({n_-} - {n_+} - {n_i}), \\
\end{array}
\end{equation}
where, $n_i$, $u_i$ and $\bar\omega=\omega_{ci}/\omega_{pi}$ ($\omega_{c_i}=eB_0/m_i$) are the ion density, velocity and plasma ion-frequencies (defined below), respectively. Also, the quantity $\sigma=T_i/T_-$ is the fractional ion-temperature and $n_\pm$ are the number densities of inertialless charged ingredients. The dimensionless set of equations, Eqs. (\ref{normal}) has been obtained using a general scaling defined below
\begin{equation}\label{nm}
\nabla \to \frac{1}{\lambda_i}\bar \nabla,\hspace{2mm}t \to \frac{{\bar t}}{{{\omega _{pi}}}},\hspace{2mm}{n} \to {n_{-0}}{\bar n}, \hspace{2mm}{\bf{u_i}} \to {c_i}{\bf{\bar u_i}},\hspace{2mm}\varphi  \to \frac{\nu}{e}\bar \varphi,
\end{equation}
where, $\omega_{pi}=\sqrt{4\pi e^2 n_0/m_i}$, ${\lambda_i} = c_i/\omega_{pi}$ and $c_i=\sqrt{\nu/m_i}$ are the characteristic plasma frequency, ion gyroradius and sound-speed the values of which along with the parameter $\nu$ will be defined later based on the charge distribution. Note that, we only consider the low-frequency ion-acoustic solitary waves (IASWs) in which $\omega_{pi}\ll\omega_{ci}$ or equivalently when the ion thermal velocity is much less than the value $\omega_{ci}/k$. The last equation in Eqs. (\ref{normal}) may be Taylor-expanded using inertial-less charged particle distributions to give
\begin{equation}\label{dist}
\Delta \varphi  = \beta  - {n_i} + {a_1}(\alpha ,\mu )\varphi  + {a_2}(\alpha ,\mu ){\varphi ^2} +  \ldots.
\end{equation}
where,
\begin{equation}\label{dist}
\alpha  = \frac{{{n_{+0}}}}{{{n_{-0}}}},\hspace{3mm}\mu=\frac{T_{-}}{T_{+}}.\\
\end{equation}
The the equilibrium charge neutrality condition is given by Poisson's relation as
\begin{equation}
\alpha  + \beta  = 1,\hspace{3mm}\beta  = \frac{{{n_{i0}}}}{{{n_{-0}}}},
\end{equation}
To find stationary localized solutions we may transform the normalized plasma equations (Eqs. (\ref{normal})) to the appropriate stretched coordinate which admits the seperation of variables permits elimination of secular terms. The transformation which leads to the desired evolution equations and the corresponding collision phase-shifts is as follows \cite{washimi, oikawa, jeffery, masa}
\begin{equation}\label{stretch}
\begin{array}{l}
\xi  = \varepsilon (kx + ly + mz - {c_\xi }t) + {\varepsilon ^2}{P_0}(\eta ,\tau ) + {\varepsilon ^3}{P_1}(\xi ,\eta ,\tau ) +  \ldots , \\
\eta  = \varepsilon (kx + ly + mz - {c_\eta }t) + {\varepsilon ^2}{Q_0}(\xi ,\tau ) + {\varepsilon ^3}{Q_1}(\xi ,\eta ,\tau ) +  \ldots , \\
\tau  = {\varepsilon ^3}t,\hspace{3mm}{c_\xi } = c,\hspace{3mm}{c_\eta } =  - c, \\
\end{array}
\end{equation}
where, the functions $P_j$ and $Q_j$ ($j=0,1,2,...$) describe the phase changes in the traveling solitary waves. These quantities will be determined later along with the wave evolution equations. It is also assumed that the interacting solitons are initially far apart and travel towards each other and described by cosine indices $(k,l,m)$ relative to a magnetic field lines with angle $\gamma$ defined through
\begin{equation}
\begin{array}{l}
m = \cos\gamma,\\
k^2 + l^2 + m^2 = 1.\\
\end{array}
\end{equation}
Expanding the dependent plasma variables around their equilibrium values through smallness $\varepsilon$ parameter, a measure of nonlinearity strength \cite{infeld}, we obtain
\begin{equation}\label{Ordering}
\left[ {\begin{array}{*{20}{c}}
{{n_i}}  \\
{\begin{array}{*{20}{c}}
{{u_i}}  \\
{{v_i}}  \\
{{w_i}}  \\
\end{array}}  \\
\varphi   \\
\end{array}} \right] = \left[ {\begin{array}{*{20}{c}}
\beta   \\
{\begin{array}{*{20}{c}}
0  \\
0  \\
0  \\
\end{array}}  \\
0  \\
\end{array}} \right] + {\varepsilon ^2}\left[ {\begin{array}{*{20}{c}}
{n_i^{(1)}}  \\
{\begin{array}{*{20}{c}}
0  \\
0  \\
{w_i^{(1)}}  \\
\end{array}}  \\
{{\varphi ^{(1)}}}  \\
\end{array}} \right] + {\varepsilon ^3}\left[ {\begin{array}{*{20}{c}}
{n_i^{(2)}}  \\
{\begin{array}{*{20}{c}}
{u_i^{(1)}}  \\
{v_i^{(1)}}  \\
{w_i^{(2)}}  \\
\end{array}}  \\
{{\varphi ^{(2)}}}  \\
\end{array}} \right] + {\varepsilon ^4}\left[ {\begin{array}{*{20}{c}}
{n_i^{(3)}}  \\
{\begin{array}{*{20}{c}}
{u_i^{(2)}}  \\
{v_i^{(2)}}  \\
{w_i^{(3)}}  \\
\end{array}}  \\
{{\varphi ^{(3)}}}  \\
\end{array}} \right] +  \ldots
\end{equation}
Stretched state equations are presented in appendix A, isolations of which in lowest-order in $\varepsilon$ leads to the following relations
\begin{equation}\label{leading}
\begin{array}{l}
c\left( { - {\partial _\xi } + {\partial _\eta }} \right)n_i^{(1)} + m\beta \left( {{\partial _\xi } + {\partial _\eta }} \right)w_i^{(1)} = 0, \\ k\beta \left( {{\partial _\xi } + {\partial _\eta }} \right){\varphi ^{(1)}} + m\sigma \left( {{\partial _\xi } + {\partial _\eta }} \right)n_i^{(1)} - \beta \bar \omega v_i^{(1)} = 0, \\ l\beta \left( {{\partial _\xi } + {\partial _\eta }} \right){\varphi ^{(1)}} + l\sigma \left( {{\partial _\xi } + {\partial _\eta }} \right)n_i^{(1)} + \beta \bar \omega u_i^{(1)} = 0, \\ c\beta \left( { - {\partial _\xi } + {\partial _\eta }} \right)w_i^{(1)} + m\beta \left( {{\partial _\xi } + {\partial _\eta }} \right){\varphi ^{(1)}} + m\sigma \left( {{\partial _\xi } + {\partial _\eta }} \right)n_i^{(1)} = 0, \\ n_i^{(1)} = {a_1}{\varphi ^{(1)}}, \\
\end{array}
\end{equation}
from which one obtains the following first-order approximations
\begin{equation}\label{Firstcomp}
\begin{array}{l}
n_i^{(1)} = a_1\left[ {\varphi ^{(1)}(\xi ,\tau ) + \varphi ^{(1)}(\eta ,\tau )} \right], \\
u_i^{(1)} =  - \frac{{l\delta }}{{\bar \omega \beta }}\left[ {{\partial _\xi}\varphi ^{(1)}(\xi ,\tau ) + {\partial _\eta }\varphi ^{(1)}(\eta ,\tau )} \right], \\
v_i^{(1)} = \frac{{k\delta }}{{\bar \omega \beta }}\left[ {{\partial _\xi }\varphi ^{(1)}(\xi ,\tau ) + {\partial _\xi }\varphi ^{(1)}(\eta ,\tau )} \right], \\
w_i^{(1)} = \frac{{ca_1}}{{m \beta }}\left[ {\varphi ^{(1)}(\xi ,\tau ) - \varphi ^{(1)}(\eta ,\tau )} \right], \\
\delta  = \beta  + a_1. \\
\end{array}
\end{equation}
Dispersion relation is, thus, given as
\begin{equation}\label{dispers}
\frac{{\beta {m^2}}}{{{c^2} - \sigma {m^2}}} = a_1,
\end{equation}
and the corresponding phase-speed $c$ is
\begin{equation}\label{speed}
c = \sqrt {\frac{\beta  +a_1 \sigma}{{a_1}}} \cos \gamma.
\end{equation}
The next higher-order step in isolation of $\varepsilon$-power lead to the second-order plasma-state approximation of the similar form as above
\begin{equation}\label{second}
\begin{array}{l}
c\left( { - {\partial _\xi } + {\partial _\eta }} \right)n_i^{(2)} + k\left( {{\partial _\xi } + {\partial _\eta }} \right)u_i^{(1)} + l\left( {{\partial _\xi } + {\partial _\eta }} \right)v_i^{(1)} + m\left( {{\partial _\xi } + {\partial _\eta }} \right)w_i^{(2)} = 0, \\
c\beta \left( { - {\partial _\xi } + {\partial _\eta }} \right)u_i^{(1)} + k\beta \left( {{\partial _\xi } + {\partial _\eta }} \right){\varphi ^{(2)}} + k\sigma \left( {{\partial _\xi } + {\partial _\eta }} \right)n_i^{(2)} - \beta \bar \omega v_i^{(2)} = 0, \\
c\beta \left( { - {\partial _\xi } + {\partial _\eta }} \right)v_i^{(1)} + l\beta \left( {{\partial _\xi } + {\partial _\eta }} \right){\varphi ^{(2)}} + l\sigma \left( {{\partial _\xi } + {\partial _\eta }} \right)n_i^{(2)} + \beta \bar \omega u_i^{(2)} = 0, \\
c\beta \left( { - {\partial _\xi } + {\partial _\eta }} \right)w_i^{(2)} + m\beta \left( {{\partial _\xi } + {\partial _\eta }} \right){\varphi ^{(2)}} + m\sigma \left( {{\partial _\xi } + {\partial _\eta }} \right)n_i^{(2)} = 0, \\
 n_i^{(2)} = {a_1}{\varphi ^{(2)}}, \\
\end{array}
\end{equation}
which immediately yield the second-order components
\begin{equation}\label{uv2}
\begin{array}{l}
n_i^{(2)} = a_1\left[ {\varphi ^{(2)}(\xi ,\tau ) + \varphi ^{(2)}(\eta ,\tau )} \right], \\
u_i^{(2)} = \frac{{ck\delta }}{{{{\bar \omega }^2}\beta }}\left[ {{\partial _\xi }\varphi ^{(2)}(\xi ,\tau ) - {\partial _\eta }\varphi ^{(2)}(\eta ,\tau )} \right] - \frac{{l\delta }}{{\bar \omega \beta }}\left[ {{\partial _{\xi \xi }}\varphi ^{(1)}(\xi ,\tau ) + {\partial _{\eta \eta }}\varphi ^{(1)}(\eta ,\tau )} \right], \\
v_i^{(2)} = \frac{{cl\delta }}{{{{\bar \omega }^2}\beta }}\left[ {{\partial _\xi }\varphi ^{(2)}(\xi ,\tau ) - {\partial _\eta }\varphi ^{(2)}(\eta ,\tau )} \right] + \frac{{k\delta }}{{\bar \omega \beta }}\left[ {{\partial _{\xi \xi }}\varphi ^{(1)}(\xi ,\tau ) + {\partial _{\eta \eta }}\varphi ^{(1)}(\eta ,\tau )} \right], \\
w_i^{(2)} = \frac{{ca_1}}{{m\beta }}\left[ {\varphi ^{(2)}(\xi ,\tau ) - \varphi ^{(2)}(\eta ,\tau )} \right]. \\
\end{array}
\end{equation}
Note that the notations ${\varphi ^{(1)}}(\xi,\tau)$ and ${\varphi ^{(1)}}(\eta,\tau)$ describe the first-order amplitude evolution and ${\varphi ^{(2)}}(\xi,\tau)$ and ${\varphi ^{(2)}}(\eta,\tau)$ denote the second-order amplitude evolution of distinct solitary excitations in the oblique directions ${\eta_ \bot }$ and ${\xi_ \bot }$ (${\eta_ \bot=-\xi_ \bot }$), respectively. For further simplicity, we will use the notations ${\varphi_\xi ^{(1)}}$ and ${\varphi_\eta ^{(1)}}$ instead of ${\varphi ^{(1)}}(\xi,\tau)$ and ${\varphi ^{(1)}}(\eta,\tau)$ in forthcoming algebra.

\section{Dynamics of Head-on Collision and Propagation}\label{shift}

The third-order approximation for density component is obtained via isolation of terms in higher-order $\varepsilon$-power. This is accomplished by solving coupled differential equations in this order, making use of dispersion relation (Eq. \ref{dispers}) and the first- and second-order plasma approximations defined in previous section
\begin{equation}\label{n3}
\begin{array}{l}
n_i ^{(3)} ={K}{N}\left[ {\frac{{\partial \varphi _\eta^{(1)}}}{{\partial \tau}} + A\varphi _\eta^{(1)}\frac{{\partial \varphi _\eta^{(1)}}}{{\partial \eta}} - B\frac{{{\partial ^3}\varphi _\eta^{(1)}}}{{\partial {\eta^3}}}} \right]\xi -
{K}{N}\left[ {\frac{{\partial \varphi _\xi^{(1)}}}{{\partial \tau}} - A\varphi _\xi^{(1)}\frac{{\partial \varphi _\xi^{(1)}}}{{\partial \xi}} + B\frac{{{\partial ^3}\varphi _\xi^{(1)}}}{{\partial {\xi^3}}}} \right]\eta +  \\
{K{E_2}}\left[ {{P_0}(\eta,\tau) - \frac{E_1}{E_2}\int {\varphi _\eta^{(1)}d\eta} } \right]\frac{{\partial \varphi _\xi^{(1)}}}{{\partial \xi}} -
{K{E_2}}\left[ {{Q_0}(\xi,\tau) - \frac{E_1}{E_2}\int {\varphi _\xi^{(1)}d\xi} } \right]\frac{{\partial \varphi _\eta^{(1)}}}{{\partial \eta}} +  \\
{K}{N}\left[ {\int {\frac{{\partial \varphi _\xi^{(1)}}}{{\partial \tau}}d\xi}  - \int {\frac{{\partial \varphi _\eta^{(1)}}}{{\partial \tau}}d\eta}}\right] - {CK}\left[ {{{(\varphi _\xi^{(1)})}^2} - {{(\varphi _\eta^{(1)})}^2}} \right] + DK\left[ {\frac{{{\partial ^2}\varphi _\xi^{(1)}}}{{\partial {\xi^2}}} - \frac{{{\partial ^2}\varphi _\eta^{(1)}}}{{\partial {\eta^2}}}} \right] + \\
F(\xi,\tau) + G(\eta,\tau), \\
\end{array}
\end{equation}
in which, $F(\xi,\tau)$ and $G(\eta,\tau)$ denote the homogenous solutions of differential equations with the other undefined coefficients in Eq. (\ref{n3}) defined as below
\begin{equation}\label{coeffs}
\begin{array}{l}
K = \frac{\beta }{{4(\beta  + {a_1}\sigma )}},\\
A = \frac{{2{a_2}{\beta ^2} + a_1^3\sigma }}{{2a_1^2\beta \sqrt {\frac{\beta }{{{a_1}}} + \sigma } }}\cos \gamma,\\
B = \frac{{\beta {{\bar \omega }^2} + ({a_1} + \beta )(\beta  + {a_1}\sigma ){{\sin }^2}\gamma }}{{2{a_1}{{\bar \omega }^2}\sqrt {\frac{\beta }{{{a_1}}} + \sigma } }}\cos \gamma,\\
N = \frac{{2a_1^2\sqrt {\frac{\beta }{{{a_1}}} + \sigma } }}{\beta }\sec \gamma,\\
E_1 = 2{a_2} - \frac{{a_1^3\sigma }}{{{\beta ^2}}},\\
E_2 = \frac{{4{a_1}(\beta  + {a_1}\sigma )}}{\beta },\\
C = {a_2} + \frac{{a_1^3\sigma }}{{2{\beta ^2}}},\\
D = \frac{{({a_1} + \beta )(\beta  + {a_1}\sigma )}}{{\beta {{\bar \omega }^2}}}{\sin ^2}\gamma  - 1,\\
\end{array}
\end{equation}
To fully determine the dynamical characteristics of the interacting solitary waves we eliminate the secular terms appearing in Eq. (\ref{n3}), which leads to two coupled differential equations for each wave, namely
\begin{equation}\label{kdv1}
\frac{{\partial \varphi _\xi^{(1)}}}{{\partial \tau}} + A\varphi _\xi^{(1)}\frac{{\partial \varphi _\xi^{(1)}}}{{\partial \xi}} - B\frac{{{\partial ^3}\varphi _\xi^{(1)}}}{{\partial {\xi^3}}} = 0,
\end{equation}
\begin{equation}\label{P0}
{P_0}(\eta,\tau) = \frac{{{E_1}}}{{{E_2}}}\int {\varphi _\eta^{(1)}d\eta},
\end{equation}
\begin{equation}\label{kdv2}
\frac{{\partial \varphi _\eta^{(1)}}}{{\partial \tau}} - A\varphi _\eta^{(1)}\frac{{\partial \varphi _\eta^{(1)}}}{{\partial \eta}} + B\frac{{{\partial ^3}\varphi _\eta^{(1)}}}{{\partial {\eta^3}}} = 0,
\end{equation}
\begin{equation}\label{Q0}
{Q_0}(\xi,\tau) = \frac{{{E_1}}}{{{E_2}}}\int {\varphi _\xi^{(1)}d\xi},
\end{equation}
On the other hand, the single-soliton solutions for Eqs. (\ref{kdv1}) and (\ref{kdv2}) require the perturbed potential components and their derivatives to vanish at infinity, i.e.
\begin{equation}\label{boundary}
\begin{array}{l}
\mathop {\lim }\limits_{\zeta \to \pm\infty } \{\varphi _\zeta^{(1)},\frac{\partial \varphi _\zeta^{(1)}}{\partial \zeta },\frac{\partial ^2\varphi _\zeta^{(1)}}{\partial \zeta
^2}\}=0,\hspace{3mm} \zeta={\xi,\eta},
\end{array}
\end{equation}
with following solutions
\begin{equation}\label{phi-x}
\begin{array}{l}
{\varphi _\xi ^{(1)} = \frac{{{\varphi _{\xi 0}}}}{{\cosh^2(\frac{{\xi  - {u_{\xi 0}}\tau }}{{{\Delta _\xi }}})}},}  \\
{{\varphi _{\xi 0}} = \frac{{3{u_{\xi 0}}}}{{{A}}},{\Delta _\xi } = {{(\frac{{4{B}}}{{{u_{\xi 0}}}})}^{\frac{1}{2}}},}  \\
\end{array}
\end{equation}
\begin{equation}\label{phi-y}
\begin{array}{l}
\varphi _\eta^{(1)} = \frac{{{\varphi _{\eta0}}}}{{\cosh^2(\frac{{\eta + {u_{\eta0}}\tau}}{{{\Delta _\eta}}})}},\\
{\varphi _{\eta0}} = \frac{{3{u_{\eta0}}}}{A},\hspace{3mm} {\Delta _\eta} = {(\frac{{4B}}{{{u_{\eta0}}}})^{\frac{1}{2}}},
\end{array}
\end{equation}
where, $\varphi _{0}$ and $\Delta$ represent the soliton amplitude and width, respectively, and $u_{0}$ is the Mach-value. The collision phase-shifts of solitary excitations are obtained using Eqs. (\ref{P0}) and (\ref{Q0}) and the KdV solutions (Eqs. (\ref{phi-x}) and (\ref{phi-y})) as
\begin{equation}\label{phase-x}
\begin{array}{l}
{P_0}(\eta,\tau) = \frac{E_1}{E_2} \varphi _{\eta0} \Delta _\eta\tanh(\frac{{\eta - {u_{\eta0}}\tau}}{{{\Delta _\eta}}}),
\end{array}
\end{equation}
\begin{equation}\label{phase-y}
\begin{array}{l}
{Q_0}(\xi,\tau) = \frac{E_1}{E_2} \varphi _{\xi0} \Delta _\xi\tanh(\frac{{\xi + {u_{\xi0}}\tau}}{{{\Delta _\xi}}}).
\end{array}
\end{equation}
The overall phase-shifts is then obtained by comparing the phases of each wave long before and after the collision
\begin{equation}
\begin{array}{l}
\Delta {P_0} = P_{post-collision}-P_{past-collision}=\\ \mathop {\lim }\limits_{\xi=0,\eta \to  + \infty } [\varepsilon ({k}x + {l}y + {m}z + {c}t)]-
\mathop {\lim }\limits_{\xi=0,\eta \to  - \infty } [\varepsilon ({k}x + {l}y + {m}z + {c}t)] , \\
\Delta {Q_0} = Q_{post-collision}-Q_{past-collision}=\\ \mathop {\lim }\limits_{\eta=0,\xi \to  + \infty } [\varepsilon ({k}x + {l}y + {m}z - {c}t)]-
\mathop {\lim }\limits_{\eta=0,\xi \to  - \infty } [\varepsilon ({k}x + {l}y + {m}z - {c}t)] , \\
\end{array}
\end{equation}
where, the quantities $\Delta {P_0}$ and $\Delta {Q_0}$ denote the overall phase-shifts of solitons $"s1"$ and $"s2"$. Finally, making use of Eqs. (\ref{phase-x}), (\ref{phase-y}) and (\ref{stretch}), leads to the following expressions for the overall phase-shift in head-on collision
\begin{equation}\label{shifts}
\begin{array}{l}
\Delta {P_0} = - {\varepsilon ^2}\left[\frac{{2{a_2}{\beta ^2} - a_1^3\sigma }}{{4{a_1}\beta (\beta  + {a_1}\sigma )}}\right]{\varphi _{\eta 0}}{\Delta _\eta }, \\
\Delta {Q_0} = {\varepsilon ^2}\left[\frac{{2{a_2}{\beta ^2} - a_1^3\sigma }}{{4{a_1}\beta (\beta  + {a_1}\sigma )}}\right]{\varphi _{\xi 0}}{\Delta _\xi }. \\
\end{array}
\end{equation}

\section{Global Features}\label{discussion1}

It is clear that the three-component model used here reduces to the simple two-component (electron-ion) case for $\beta=1$ and to the cold-ion case for $\sigma=0$. It is also remarked from KdV coefficients, Eqs. (\ref{coeffs}), that in general the solitary wave amplitude only depends on the angle of the ambient field and is independent of its strength while the soliton width depends on both parameters. Close inspection reveals that the soliton width reaches a maximum value as the field angle varies in the range $0<\gamma<\pi/2$. The value of the maximum soliton width in this range is given by the following analytical expression
\begin{equation}\label{gamma}
{\gamma _m} = \arccos \left[ {\sqrt {\frac{1}{3}\left( {1 + \frac{{\beta {{\bar \omega }^2}}}{{(\beta  + {a_1})(\beta  + {a_1}\sigma )}}} \right)} } \right],
\end{equation}
and, at the cold-ion limit we obtain
\begin{equation}\label{gamma}
{\gamma _m} = \arccos {\left[ {\frac{1}{3}\left( {1 + \frac{{{{\bar \omega }^2}}}{{\beta  + {a_1}}}} \right)} \right]^{\frac{1}{2}}}.
\end{equation}
It is easily observed that, in general, the corresponding value of $\gamma_m$ can not exceed the limiting value of $\gamma\simeq54.74^\circ$. On the other hand, evaluation of the collision phase-shift given in Eq. (\ref{shifts}) reveals that, when the ions are warm ($\sigma\neq 0$) for some set of critical parameters the phase-shift vanishes and changes the sign. For instance, the corresponding critical fractional ion-temperature for which the phase-shift vanishes is given by $\sigma_{cr}=2a_2\beta^2/a_1^{3}$. It is interesting that these critical values are completely independent of the strength of magnetic field, $\bar\omega$ or its angle relative to the direction of wave propagation, $\gamma$.

\section{Special Cases}\label{discussion2}

The preceding arguments are valid for all magnetized three-component plasma with one positive inertial and two inertial-less ingredients with opposite charges and can be evaluated if the coefficients $a_1$ and $a_2$ are known. It can be remarked that the general calculation used above may be easily extended to magnetized plasmas containing extra immobile background species with either charges. However, in order to investigate the effect of various plasma parameters on propagation and head-on collision of ion-acoustic solitary waves, we presented in subsections below some specific cases which might be of interest. We also emphasize on the Kappa (Tsallis) distribution which might be of great importance in astrophysics.

\subsection{Maxwell-Boltzmann Distribution}

To start with we consider the most simple case of Boltzmann electron/positron. For Maxwell-Boltzmann electron/positron distribution in classical electron-positron-ion plasma, we have $n_e=e^{\varphi}$ and $n_p=(1-\beta)e^{-\mu\varphi}$ with $\sigma=T_i/T_{e}$, $\mu=T_p/T_e$ and $\nu=k_B T_e$. Then we derive ${a_1} = 1 + \mu (1 - \beta )$ and ${a_2} = (1 - {\mu ^2}(1 - \beta ))/2$. Characteristics of solitary ion-acoustic waves in such plasma has been considered in Ref. \cite{akbari5} which is consistent with the general features mentioned in previous section.

\subsection{Zero-Temperature Thomas-Fermi Distribution}

For the Thomas-Fermi electron/positron distribution in completely degenerate electron-positron-ion plasma, we have $n_e={(1 + \varphi )^{3/2}}$ and $n_p=(1-\beta){(1 + \mu\varphi )^{3/2}}$ with $\sigma=T_i/T_{Fe}$, $\mu=T_{Fp}/T_{Fe}=(1-\beta)^{-2/3}$ for nonrelativistic distribution and $\mu=T_{Fp}/T_{Fe}=(1-\beta)^{-1/3}$ for ultrarelativistic electron/positron distribution and $\nu=k_B T_{Fe}$. Then we derive ${a_1} = 3(1 + \mu (1 - \beta ))/2$ and ${a_2} = 3(1 - {\mu ^2}(1 - \beta ))/8$. It is observed that, in this case the value of $\sigma_{cr}$ (defined above) is $1/9$-times that of the Maxwell-Boltzmann electron-positron-ion plasma. Dynamics of solitary ion-acoustic waves and their head-on collision in this plasma with both ultrarelativistic and nonrelativistic electrons and positrons has been considered in Ref. \cite{akbari6} and agrees with the general features mentioned above. Also, similar features can be observed for Fermi-Dirac electron-positron-ion plasma considered in Ref. \cite{akbari7}.

\subsection{Partially Degenerate Thomas-Fermi Distribution}

For the Thomas-Fermi electron/positron distribution in partially degenerate electron-ion plasma, we have $n_e={(1 + \varphi )^{3/2}}+T_e^2(1+\varphi)^{-1/2}$ and $\beta=1/(1+T_e^2)$ with $\sigma=T_i/T_{e}$, $\mu=0$ and $\nu=k_B T_{e}$. Then we derive ${a_1} = (3 + {T_e^2})/2$ and ${a_2} = 3(1 - {T_e^2})/8$. It is observed that, in this case the value of $\sigma_{cr}$ solely depends on the electron temperature, $T_e$.

\subsection{Tsallis Distribution}

\textbf{Tsallis \cite{tsallis} in 1988 has proposed a generalization of Boltzmann-Gibbs statistics based on multi-fractal concept of probability. The observation based on interstellar plasma velocity distribution measurments confirms that non-Maxwellian distributions are common in the solar wind and in the planetary magnetospheres where the velocity distribution have a "Tsallis-like" power-law tail at high energies \cite{kohl}. The Tsallis velocity distribution is a convenient extension of the well-known Boltzmann velocity distribution in the sense that it reduces to the ordinary Boltzmann distribution in the limiting case of the spectral index \cite{silva}. However, there are various equivalent versions to characterize the q-nonextensive (Tsallis) velocity distributions usually employed in the literature. Here, we consider the distribution adopted in the Refs. \cite{liu, zipeng1, zipeng2} for inertialless ingredients keeping non-Tsallis distribution for ions. The reason is to avoid the Landau damping by keeping the fractional ion to inertialless-particle temperature $\sigma$ much smaller than unity. Hence, for the superthermal electrons/positrons in electron-positron-ion plasma, we have for the spectral index values $q > 1$ \cite{wady} ${n_e} = {[1 + (q - 1)\varphi ]^{(q + 1)/2(q-1)}}$ and ${n_p} = \alpha{[1 + (q - 1)\mu\varphi ]^{(q + 1)/2(q-1)}}$ with $\sigma=T_i/T_{e}$, $\mu=T_p/T_e$, $\alpha=1-\beta$ and $\nu=k_B T_e$. Hence, we derive ${a_1} = (1 + q)(1 + \alpha \mu )/2$ and ${a_2} = (3-q)(1 + q)(1 + \alpha {\mu ^2})/8$. It can be easily verified that the special case of $q=1$ corresponds to the well-known Maxwell-Boltzmann distribution given in Sec. \ref{mb}. Let us now consider in detail the dynamics of propagation and head-on collision of general Tsallis-distributed magnetized plasma which may be encountered in various astrophysical situations.}

\textbf{Figure 1 present the variations of the soliton amplitude with respect to fractional positron to electron temperature $\mu$ for different fractional ion number-density $\beta$. It is noted from Fig. 1(a) that for some set of values of $\mu$, $\beta$ and $\sigma$ the soliton changes from compressive to rarefactive and viceversa. It is observed from Fig. 1(b) that a critical angle $\gamma=\pi/2$ and critical fractional positron temperature exist which define the shape (brightness/darkness) of the solitons.}

\textbf{On the other hand, Fig. 2 presents the variations in the soliton width with respect to different plasma parameters. It is clearly remarked from Fig 2(a)that the soliton width decreases with increase of $\mu$ while it increases with increase of the fractional ion temperature, $\sigma$. Also, Fig. 2(b) indicates that the soliton width has a maximum value in the range $0<\gamma<\pi/2$ regarding the angle of the propagation with respect to that of the field, in agreement with the general features mentioned in previous section. It is further remarked from Fig. 2(b) that the increase in the value of spectral index, $d$ leads to decrease in the soliton width for angles of the ambient magnetic field.}

\textbf{The head-on collision phase-shift and its variations with respect to various plasma parameters are given in Fig. 3. It is clearly remarked from Fig. 3(a) that the sign of the collision phase-shift can be positive or negative, in general, depending on the chosen plasma parameter set. A positive phase-shift indicates that the post-collision parts of the soliton moves ahead of the initial trajectory, whereas, a negative phase-shift denotes that the post-collision parts of the soliton lags behind the initial trajectory \cite{akbari8}. It is also confirmed from Fig. 3(b)that the spectral index has significant effect on the value and the sign of the collision phase-shift. Figures 3(b) and 3(d) reveal that the sign and the value of the collision phase-shift with varied strength of the ambient field can be much different below and above a critical plasma parameter, let say in this case, $\beta_{cr}$.}

\section{Summary}\label{conclusion}

Using an extended multiple scales technique, which includes the phase-shift variations, we showed that in a magnetized plasmas with diverse inertial-less ingredient distributions common rules apply on propagation and head-on collisions of ion-acoustic waves. In general, the soliton amplitude is independent of magnetic field strength but strictly depends on its angle regarding the propagation direction. It was also shown that the type (dark or bright) of the solitons are defined through the critical angle $\gamma=\pi/2$. Moreover, it was shown that one or more critical plasma parameter values may exist defining the sign of collision phase shift. Current study may be applied to a wide variety of magnetized plasmas including those containing background dust ingredients or two-temperature inertial-less particles.

\appendix

\section{Stretched Plasma Equations}
\begin{equation}\label{strain1}
\begin{array}{l}
{\varepsilon ^2}\frac{{\partial {n_i}}}{{\partial \tau }} - c\frac{{\partial {n_i}}}{{\partial \xi }} - {\varepsilon ^2}c\frac{{\partial {Q_0}}}{{\partial \xi }}\frac{{\partial {n_i}}}{{\partial \eta }} + c\frac{{\partial {n_i}}}{{\partial \eta }} + {\varepsilon ^2}c\frac{{\partial {P_0}}}{{\partial \eta }}\frac{{\partial {n_i}}}{{\partial \xi }} + k\frac{{\partial {n_i}{u_i}}}{{\partial \xi }} +\\ {\varepsilon ^2}k\frac{{\partial {Q_0}}}{{\partial \xi }}\frac{{\partial {n_i}{u_i}}}{{\partial \eta }} + k\frac{{\partial {n_i}{u_i}}}{{\partial \eta }} + {\varepsilon ^2}k\frac{{\partial {P_0}}}{{\partial \eta }}\frac{{\partial {n_i}{u_i}}}{{\partial \xi }} + l\frac{{\partial {n_i}{v_i}}}{{\partial \xi }} + {\varepsilon ^2}l\frac{{\partial {Q_0}}}{{\partial \xi }}\frac{{\partial {n_i}{v_i}}}{{\partial \eta }} +\\ l\frac{{\partial {n_i}{v_i}}}{{\partial \eta }} + {\varepsilon ^2}l\frac{{\partial {P_0}}}{{\partial \eta }}\frac{{\partial {n_i}{v_i}}}{{\partial \xi }} + m\frac{{\partial {n_i}{w_i}}}{{\partial \xi }} + {\varepsilon ^2}m\frac{{\partial {Q_0}}}{{\partial \xi }}\frac{{\partial {n_i}{w_i}}}{{\partial \eta }} + m\frac{{\partial {n_i}{w_i}}}{{\partial \eta }} +\\ {\varepsilon ^2}m\frac{{\partial {P_0}}}{{\partial \eta }}\frac{{\partial {n_i}{w_i}}}{{\partial \xi }} + \ldots  = 0, \\
\end{array}
\end{equation}
\begin{equation}\label{strain2}
\begin{array}{l}
{\varepsilon ^2}\frac{{\partial {u_i}}}{{\partial \tau }} - c\frac{{\partial {u_i}}}{{\partial \xi }} - {\varepsilon ^2}c\frac{{\partial {Q_0}}}{{\partial \xi }}\frac{{\partial {u_i}}}{{\partial \eta }} + c\frac{{\partial {u_i}}}{{\partial \eta }} + {\varepsilon ^2}c\frac{{\partial {P_0}}}{{\partial \eta }}\frac{{\partial {u_i}}}{{\partial \xi }} + k{u_i}\frac{{\partial {u_i}}}{{\partial \xi }} +  \\
{\varepsilon ^2}k{u_i}\frac{{\partial {Q_0}}}{{\partial \xi }}\frac{{\partial {u_i}}}{{\partial \eta }} + k{u_i}\frac{{\partial {u_i}}}{{\partial \eta }} + {\varepsilon ^2}k{u_i}\frac{{\partial {P_0}}}{{\partial \eta }}\frac{{\partial {u_i}}}{{\partial \xi }} + l{v_i}\frac{{\partial {u_i}}}{{\partial \xi }} + {\varepsilon ^2}l{v_i}\frac{{\partial {Q_0}}}{{\partial \xi }}\frac{{\partial {u_i}}}{{\partial \eta }} +  \\
l{v_i}\frac{{\partial {u_i}}}{{\partial \eta }} + {\varepsilon ^2}l{v_i}\frac{{\partial {P_0}}}{{\partial \eta }}\frac{{\partial {u_i}}}{{\partial \xi }} + m{w_i}\frac{{\partial {u_i}}}{{\partial \xi }} + {\varepsilon ^2}m{w_i}\frac{{\partial {Q_0}}}{{\partial \xi }}\frac{{\partial {u_i}}}{{\partial \eta }} + m{w_i}\frac{{\partial {u_i}}}{{\partial \eta }} +  \\
{\varepsilon ^2}m{w_i}\frac{{\partial {P_0}}}{{\partial \eta }}\frac{{\partial {u_i}}}{{\partial \xi }} + k\frac{{\partial \varphi }}{{\partial \xi }} + {\varepsilon ^2}k\frac{{\partial {Q_0}}}{{\partial \xi }}\frac{{\partial \varphi }}{{\partial \eta }} + k\frac{{\partial \varphi }}{{\partial \eta }} + {\varepsilon ^2}k\frac{{\partial {P_0}}}{{\partial \eta }}\frac{{\partial \varphi }}{{\partial \xi }} +  \\
k\frac{\sigma}{n_i}\frac{{\partial {n_i}}}{{\partial \xi }} + {\varepsilon ^2}k\frac{\sigma}{n_i}\frac{{\partial {Q_0}}}{{\partial \xi }}\frac{{\partial {n_i}}}{{\partial \eta }} + k\frac{\sigma}{n_i}\frac{{\partial {n_i}}}{{\partial \eta }} + {\varepsilon ^2}k\frac{\sigma}{n_i}\frac{{\partial {P_0}}}{{\partial \eta }}\frac{{\partial {n_i}}}{{\partial \xi }} -\\ \frac{{\bar \omega {v_i}}}{\varepsilon } + \ldots  = 0, \\
\end{array}
\end{equation}
\begin{equation}\label{strain3}
\begin{array}{l}
{\varepsilon ^2}\frac{{\partial {v_i}}}{{\partial \tau }} - c\frac{{\partial {v_i}}}{{\partial \xi }} - {\varepsilon ^2}c\frac{{\partial {Q_0}}}{{\partial \xi }}\frac{{\partial {v_i}}}{{\partial \eta }} + c\frac{{\partial {v_i}}}{{\partial \eta }} + {\varepsilon ^2}c\frac{{\partial {P_0}}}{{\partial \eta }}\frac{{\partial {v_i}}}{{\partial \xi }} + k{u_i}\frac{{\partial {v_i}}}{{\partial \xi }} +  \\
{\varepsilon ^2}k{u_i}\frac{{\partial {Q_0}}}{{\partial \xi }}\frac{{\partial {v_i}}}{{\partial \eta }} + k{u_i}\frac{{\partial {v_i}}}{{\partial \eta }} + {\varepsilon ^2}k{u_i}\frac{{\partial {P_0}}}{{\partial \eta }}\frac{{\partial {v_i}}}{{\partial \xi }} + l{v_i}\frac{{\partial {v_i}}}{{\partial \xi }} + {\varepsilon ^2}l{v_i}\frac{{\partial {Q_0}}}{{\partial \xi }}\frac{{\partial {v_i}}}{{\partial \eta }} +  \\
l{v_i}\frac{{\partial {v_i}}}{{\partial \eta }} + {\varepsilon ^2}l{v_i}\frac{{\partial {P_0}}}{{\partial \eta }}\frac{{\partial {v_i}}}{{\partial \xi }} + m{w_i}\frac{{\partial {v_i}}}{{\partial \xi }} + {\varepsilon ^2}m{w_i}\frac{{\partial {Q_0}}}{{\partial \xi }}\frac{{\partial {v_i}}}{{\partial \eta }} + m{w_i}\frac{{\partial {v_i}}}{{\partial \eta }} +  \\
{\varepsilon ^2}m{w_i}\frac{{\partial {P_0}}}{{\partial \eta }}\frac{{\partial {v_i}}}{{\partial \xi }} + l\frac{{\partial \varphi }}{{\partial \xi }} + {\varepsilon ^2}l\frac{{\partial {Q_0}}}{{\partial \xi }}\frac{{\partial \varphi }}{{\partial \eta }} + l\frac{{\partial \varphi }}{{\partial \eta }} + {\varepsilon ^2}l\frac{{\partial {P_0}}}{{\partial \eta }}\frac{{\partial \varphi }}{{\partial \xi }} +  \\
l\frac{\sigma}{n_i}\frac{{\partial {n_i}}}{{\partial \xi }} + {\varepsilon ^2}l\frac{\sigma}{n_i}\frac{{\partial {Q_0}}}{{\partial \xi }}\frac{{\partial {n_i}}}{{\partial \eta }} + l\frac{\sigma}{n_i}\frac{{\partial {n_i}}}{{\partial \eta }} + {\varepsilon ^2}l\frac{\sigma}{n_i}\frac{{\partial {P_0}}}{{\partial \eta }}\frac{{\partial {n_i}}}{{\partial \xi }} +\\ \frac{{\bar \omega {u_i}}}{\varepsilon } + \ldots  = 0, \\
\end{array}
\end{equation}
\begin{equation}\label{strain4}
\begin{array}{l}
{\varepsilon ^2}\frac{{\partial {w_i}}}{{\partial \tau }} - c\frac{{\partial {w_i}}}{{\partial \xi }} - {\varepsilon ^2}c\frac{{\partial {Q_0}}}{{\partial \xi }}\frac{{\partial {w_i}}}{{\partial \eta }} + c\frac{{\partial {w_i}}}{{\partial \eta }} + {\varepsilon ^2}c\frac{{\partial {P_0}}}{{\partial \eta }}\frac{{\partial {w_i}}}{{\partial \xi }} + k{u_i}\frac{{\partial {w_i}}}{{\partial \xi }} +  \\
{\varepsilon ^2}k{u_i}\frac{{\partial {Q_0}}}{{\partial \xi }}\frac{{\partial {w_i}}}{{\partial \eta }} + k{u_i}\frac{{\partial {w_i}}}{{\partial \eta }} + {\varepsilon ^2}k{u_i}\frac{{\partial {P_0}}}{{\partial \eta }}\frac{{\partial {w_i}}}{{\partial \xi }} + l{v_i}\frac{{\partial {w_i}}}{{\partial \xi }} + {\varepsilon ^2}l{v_i}\frac{{\partial {Q_0}}}{{\partial \xi }}\frac{{\partial {w_i}}}{{\partial \eta }} +  \\
l{v_i}\frac{{\partial {w_i}}}{{\partial \eta }} + {\varepsilon ^2}l{v_i}\frac{{\partial {P_0}}}{{\partial \eta }}\frac{{\partial {w_i}}}{{\partial \xi }} + m{w_i}\frac{{\partial {w_i}}}{{\partial \xi }} + {\varepsilon ^2}m{w_i}\frac{{\partial {Q_0}}}{{\partial \xi }}\frac{{\partial {w_i}}}{{\partial \eta }} + m{w_i}\frac{{\partial {w_i}}}{{\partial \eta }} +  \\
{\varepsilon ^2}m{w_i}\frac{{\partial {P_0}}}{{\partial \eta }}\frac{{\partial {w_i}}}{{\partial \xi }} + m\frac{{\partial \varphi }}{{\partial \xi }} + {\varepsilon ^2}m\frac{{\partial {Q_0}}}{{\partial \xi }}\frac{{\partial \varphi }}{{\partial \eta }} + m\frac{{\partial \varphi }}{{\partial \eta }} + {\varepsilon ^2}m\frac{{\partial {P_0}}}{{\partial \eta }}\frac{{\partial \varphi }}{{\partial \xi }} +  \\
m\frac{\sigma}{n_i}\frac{{\partial {n_i}}}{{\partial \xi }} + {\varepsilon ^2}m\frac{\sigma}{n_i}\frac{{\partial {Q_0}}}{{\partial \xi }}\frac{{\partial {n_i}}}{{\partial \eta }} + m\frac{\sigma}{n_i}\frac{{\partial {n_i}}}{{\partial \eta }} + {\varepsilon ^2}m\frac{\sigma}{n_i}\frac{{\partial {P_0}}}{{\partial \eta }}\frac{{\partial {n_i}}}{{\partial \xi }} +\\ \ldots  = 0, \\
\end{array}
\end{equation}
\begin{equation}\label{strain5}
\begin{array}{l}
{\varepsilon ^2}\left[ {\frac{{{\partial ^2}\varphi }}{{\partial {\xi ^2}}} + \frac{{{\partial ^2}\varphi }}{{\partial {\eta ^2}}}} \right] - \left[\beta - {n_i} + a_1\varphi  + a_2{\varphi ^2}\right] +  \\
\ldots  = 0. \\
\end{array}
\end{equation}

\newpage

\textbf{FIGURE CAPTIONS}

\bigskip

Figure 1

\bigskip

(Color online) The variation of soliton amplitude with respect to various fractional plasma parameters, such as fractional positron to electron temperature $\mu$ for different fractional plasma parameters such as the relative ion-temperature $\sigma$, the relative ion number-density $\beta$, the spectral-index $q$ and the ambient magnetic-field angle $\gamma$, while the other parameters are kept fixed. The values of $\varepsilon=0.1$ and $u_{\xi,0}=u_{\eta,0}=0.1$ are used for all plots in this figure. The dash sizes in all plots are appropriately related to the values of varied parameter.

\bigskip

Figure 2

\bigskip

(Color online) The variation of soliton width with respect to various fractional plasma parameters, such as fractional positron to electron temperature $\mu$ for different fractional plasma parameters such as the relative ion-temperature $\sigma$, the relative ion number-density $\beta$, the normalized magnetic field strength $\bar\omega$ and the ambient magnetic-field angle $\gamma$, while the other parameters are kept fixed. The values of $\varepsilon=0.1$ and $u_{\xi,0}=u_{\eta,0}=0.1$ are used for all plots in this figure. The dash sizes in all plots are appropriately related to the values of varied parameter.

\bigskip
\bigskip

Figure 3

\bigskip

(Color online) The variation of collision phase-shift with respect to various fractional plasma parameters, such as fractional positron to electron temperature $\mu$ for different fractional plasma parameters such as the relative ion-temperature $\sigma$, the relative ion number-density $\beta$, the normalized magnetic field strength $\bar\omega$, the spectral-index $q$ and the ambient magnetic-field angle $\gamma$, while the other parameters are kept fixed. The values of $\varepsilon=0.1$ and $u_{\xi,0}=u_{\eta,0}=0.1$ are used for all plots in this figure. The dash sizes in all plots are appropriately related to the values of varied parameter.

\bigskip

\newpage

\begin{figure}[ptb]\label{Figure1}
\includegraphics[scale=.6]{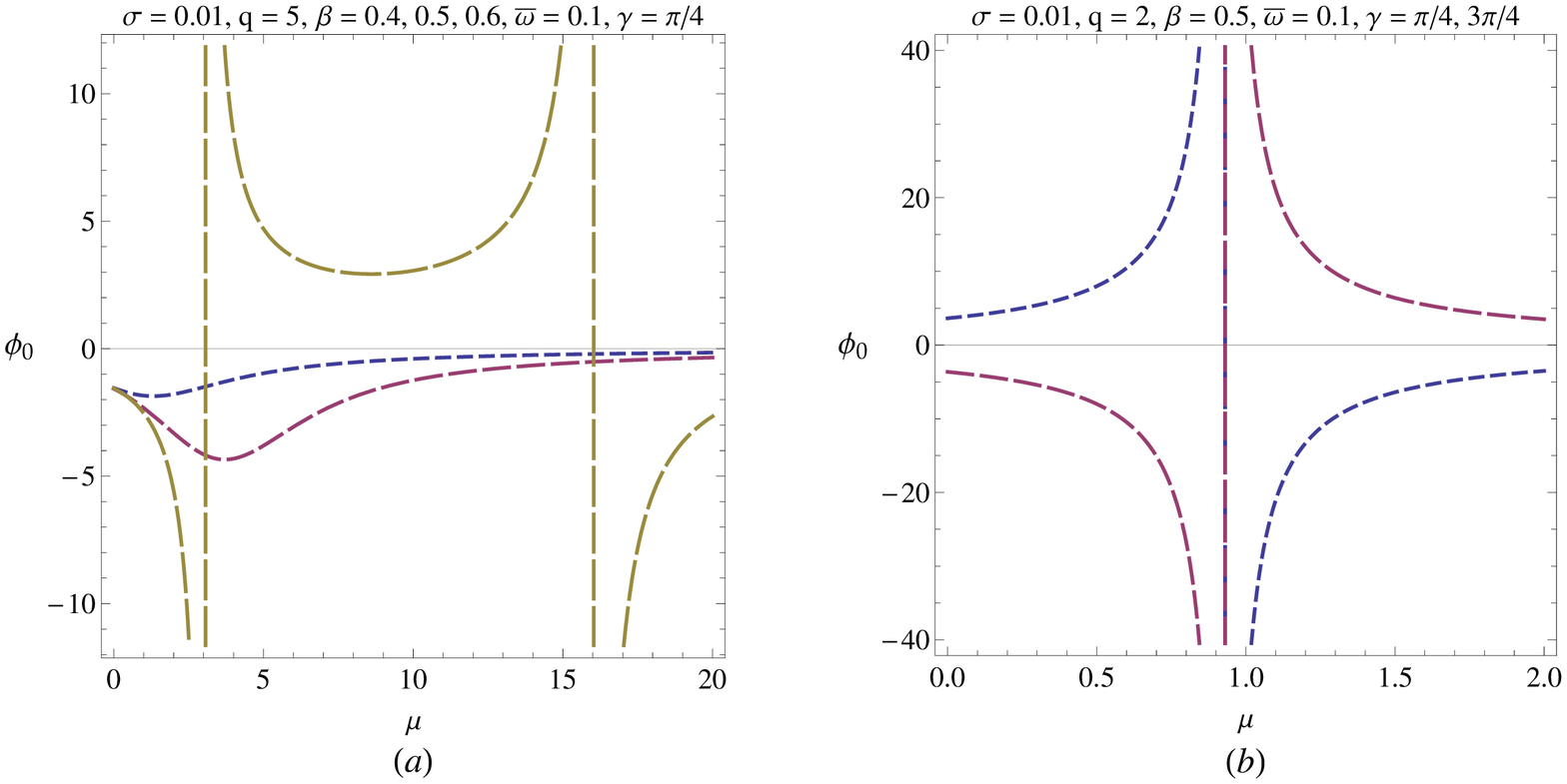}\caption{}
\end{figure}

\newpage

\begin{figure}[ptb]\label{Figure2}
\includegraphics[scale=.6]{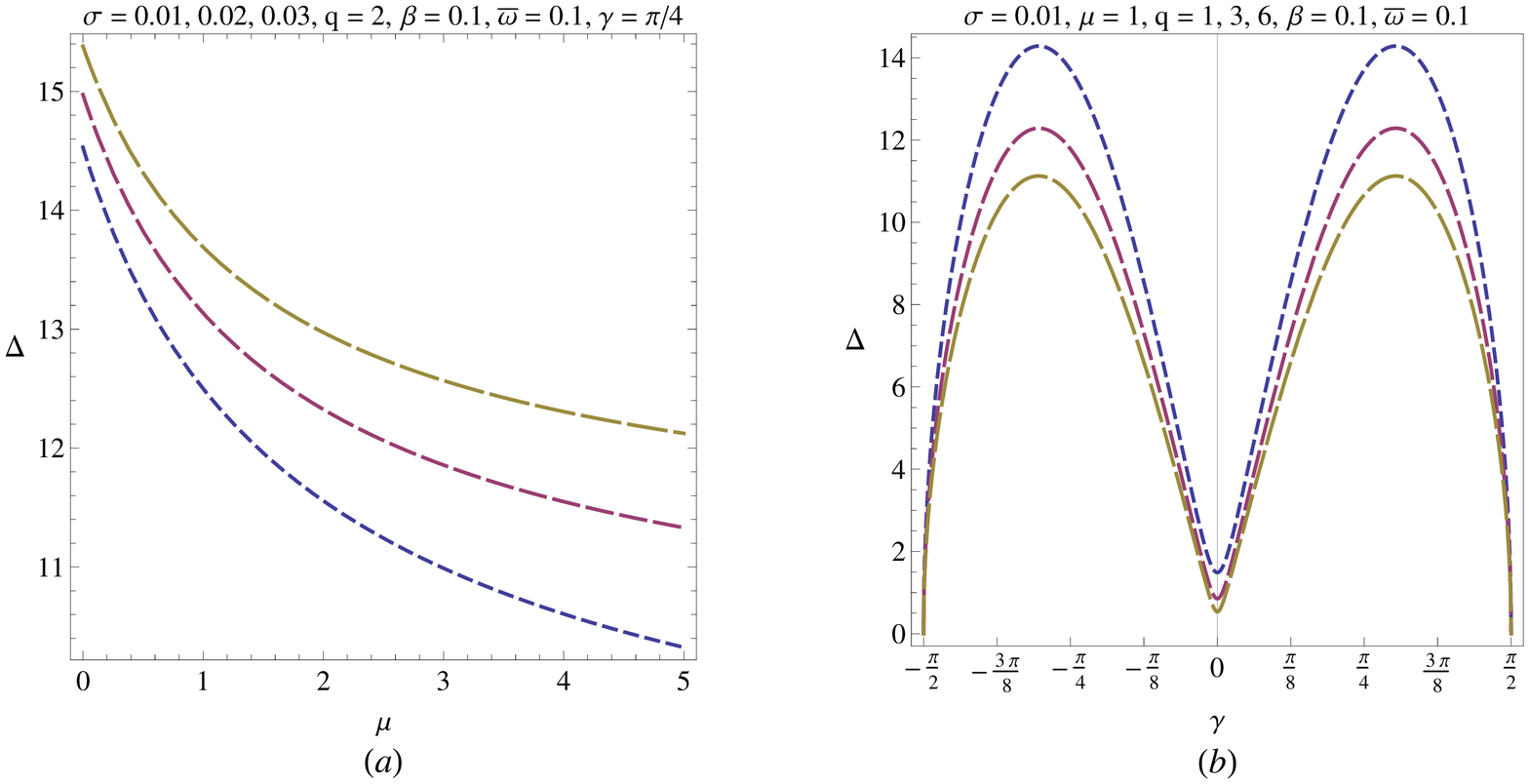}\caption{}
\end{figure}

\newpage

\newpage

\begin{figure}[ptb]\label{Figure3}
\includegraphics[scale=.6]{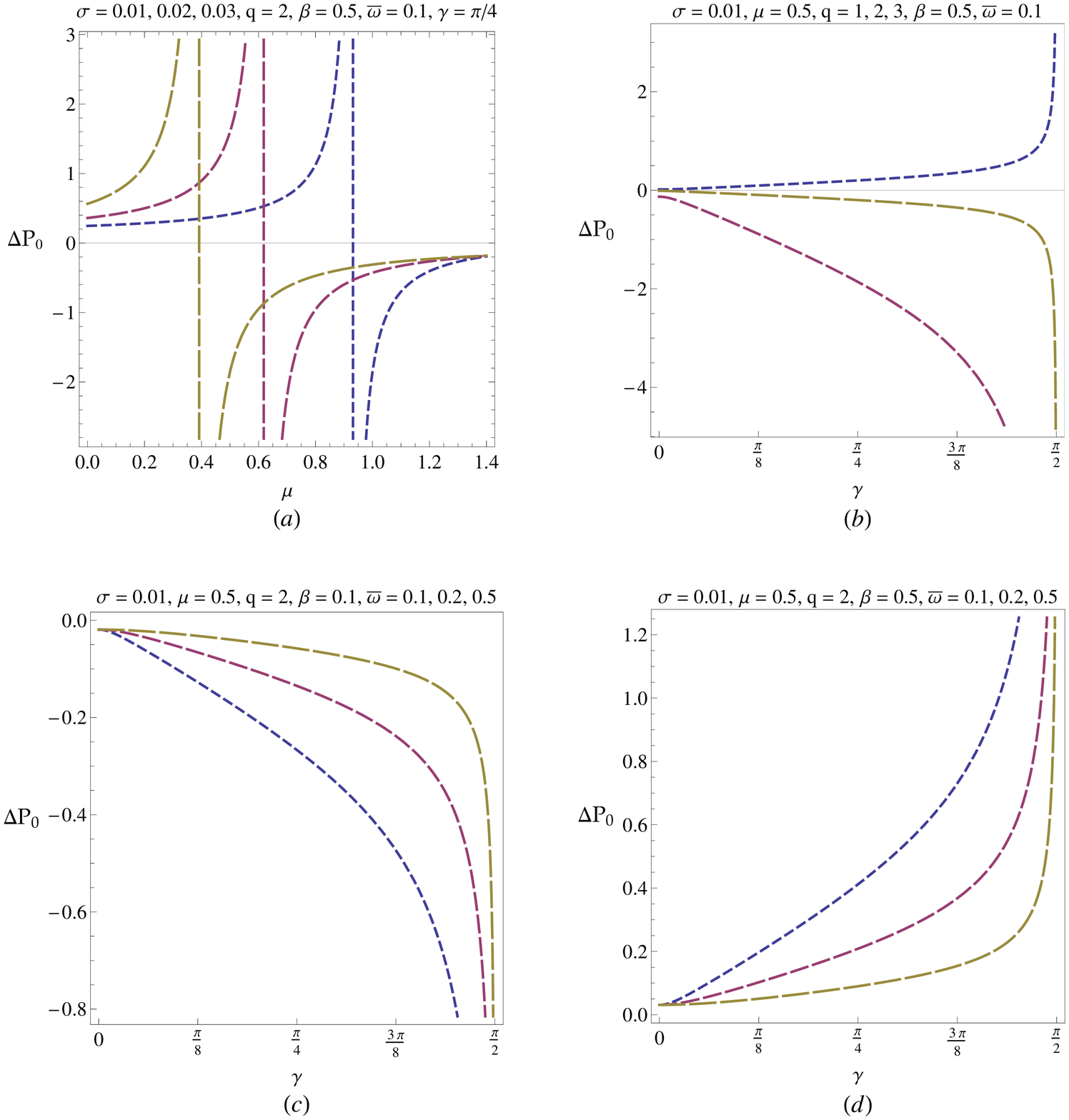}\caption{}
\end{figure}

\newpage

\end{document}